# Structural and Magnetic Properties of V-Ti-Si Alloy Superconductors

Asi Khandelwal[1,2, a)], L.S. Sharath Chandra[1,2], Archna Sagdeo[2,3], Rashmi Singh[4] and M.K. Chattopadhyay[1,2]

[1]*Free Electron Laser Utilization Laboratory, Raja Ramanna Centre for Advanced Technology, Indore, Madhya Pradesh - 452 013, India*
[2]*Homi Bhabha National Institute, Training School Complex, Anushakti Nagar, Mumbai 400 094, India*
[3]*Hard X-ray Applications Laboratory, Accelerator Physics and Synchrotrons Utilization Division, Raja Ramanna Centre for Advanced Technology, Indore, Madhya Pradesh - 452 013, India*
[4]*Nano-Functional Materials Laboratory, Laser and Functional Materials Division, Raja Ramanna Centre for Advanced Technology, Indore, Madhya Pradesh - 452 013, India*

[a)]Corresponding author: asikhandelwal1503@gmail.com

**Abstract.** The structural and magnetic properties of the as-cast and annealed $V_{0.6-x}Si_xTi_{0.4}$ (x = 0, 0.05, 0.10, 0.15) alloy superconductors are reported here. It is found that addition of silicon to the V-Ti alloys results in eutectic precipitation of $Ti_5Si_3$-phase in the body centred cubic (bcc) β-V-Ti matrix. In the as-cast $V_{0.6-x}Si_xTi_{0.4}$ alloys, the superconducting transition temperature ($T_C$) changes non-monotonically with increasing silicon content whereas after annealing, it is about 7.7 K for all the alloys. On the other hand, the upper critical field decreases and the coherence length increases after annealing in the x = 0.10 alloy. The variations in the superconducting properties in the alloys are related to the solubility of 6 at.% Si in the $V_{0.60}Ti_{0.40}$ alloy and the vanadium enrichment in the β matrix due to the precipitation of $Ti_5Si_3$ phase.

## INTRODUCTION

The V-Ti alloys have shown potential for high field applications, especially in the neutron irradiation environment [1]. Additionally, the $V_{1-x}Ti_x$ alloys are highly machinable and ductile [1, 2]. The addition of suitable β stabilizers to titanium significantly enhances the mechanical strength of the Ti alloys by effectively controlling the decomposition of the β phase into the hcp (hexagonal close-packed) α phase during the heat treatment process [3]. Adding small amount of silicon (usually less than ~1 at. %) is a well-established technique to enhance the strength of α (hcp)/ β (bcc) and martensitic Ti alloys. Addition of silicon to titanium results in the precipitation of different structural phases such as $Ti_3Si$ and $Ti_5Si_3$, depending on the composition [4-6]. On the other hand, the addition of small amount of silicon to vanadium produces a eutectic microstructure with a eutectic composition of 12.5 at.% silicon [4]. It has been reported that the dislocations pile up at the grain boundaries in the V-Si alloys [4]. The V-Ti-Si ternary alloy forms in multiple phases depending on the composition [7]. At low Si concentrations in the $V_{0.60}Ti_{0.40}$ alloy, a eutectic microstructure is formed consisting of $(V_{1-x}Ti_x)_ySi_z$ and $V_{1-x}Ti_x$ phases [7-10]. While amorphous titanium rich V-Ti-Si alloys are not superconducting, annealing is found to induce superconductivity above liquid He temperatures (4.2 K) [11]. However, a detailed study on the superconductivity in the vanadium rich V-Ti-Si alloys is not reported in literature. In this direction, we study the structural and magnetic properties of the V-Ti-Si alloy superconductors and also show that the precipitation of $Ti_5Si_3$ phase influences the superconducting properties of V-Ti-Si alloys.

## EXPERIMENTAL DETAILS

The polycrystalline $V_{0.6-x}Si_xTi_{0.4}$ (x = 0, 0.05, 0.10, 0.15) samples were synthesized by arc-melting, under 99.999% pure Ar atmosphere, with high purity (> 99.8%) starting materials, wherein the samples were homogenized by flipping-and-melting four times. Portions of these ingots were wrapped in tantalum foil and sealed in quartz ampules before annealing at 650˚C for 24 hours. The x-ray diffraction (XRD) studies were performed using a D8 Advance Diffractometer (Bruker GmbH, Germany). After polishing and etching a portion of the sample, high-resolution images were obtained using a scanning electron microscope (SEM, Carl Zeiss, Germany). Additionally, an energy dispersive analysis of x-ray (EDAX) set-up attached to the SEM was used to determine the composition at various locations on the sample. A Superconducting Quantum Interference Device based Vibrating Sample

Magnetometer (MPMS-3 SQUID VSM, Quantum Design, USA) was used for doing the magnetization (M) measurements.

## RESULTS AND DISCUSSIONS

Figure 1 shows the X-ray diffraction (XRD) patterns of the (a) as-cast and (b) annealed $V_{0.6-x}Si_xTi_{0.4}$ alloys. There is no peak corresponding to the $Ti_5Si_3$-phase in the XRD patterns of the x = 0.05 alloy. For the x > 0.05 alloys, reflections corresponding to the β-phase with a body centred cubic (bcc) structure (space group: $Im\bar{3}m$) and a secondary hexagonal $Ti_5Si_3$ phase with $P6_3/mmc$ symmetry are seen. The XRD patterns indicate no change in phases after annealing. The major peaks of both the phases are indexed using PowderCell software [12]. The lattice parameters of both the phases are estimated from the peak positions, and it is observed that the lattice parameter of the bcc phase reduces with Si addition and further decreases after annealing. This indicates vanadium enrichment in the β-phase after the addition of Si as well as after annealing. In contrast, the lattice parameter of the $Ti_5Si_3$ phase reduces with Si addition but remains almost invariant after annealing.

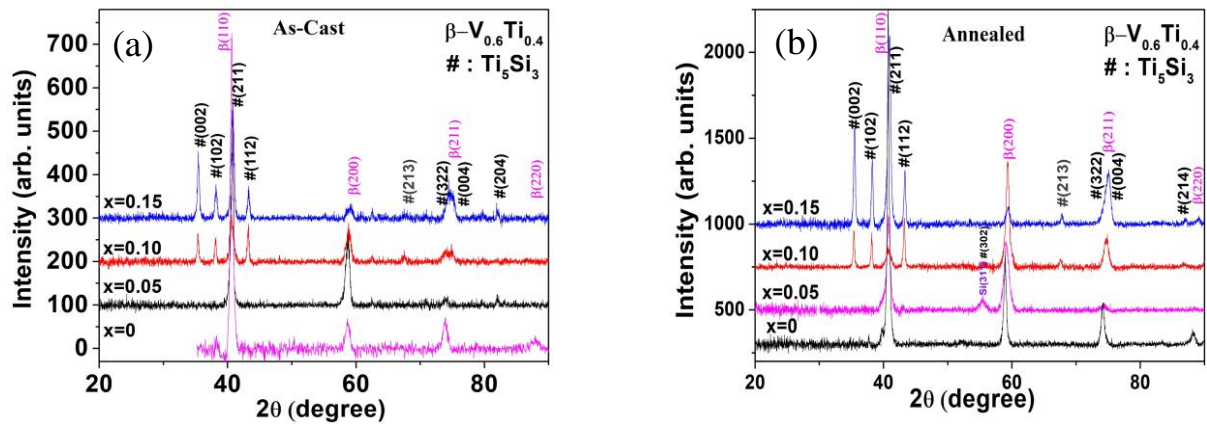

**FIGURE 1.** XRD patterns of $V_{0.6-x}Si_xTi_{0.4}$ alloys at room temperature. All the major peaks are indexed with β-V-Ti and $Ti_5Si_3$ phases.

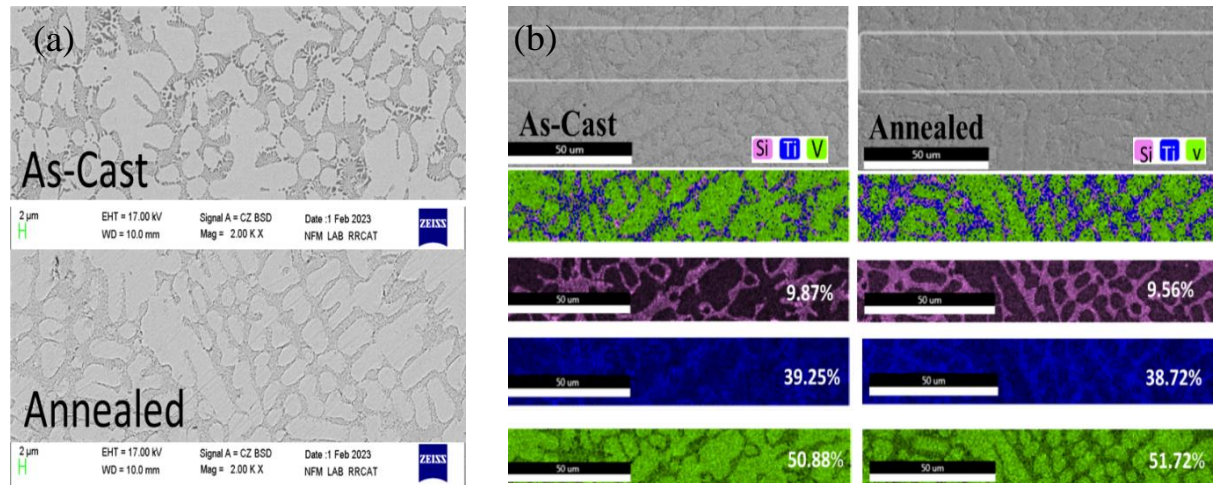

**FIGURE 2.** (a) SEM images of the polished surface of as-cast and annealed $V_{0.5}Si_{0.1}Ti_{0.4}$ alloys after etching. (b) Elemental analysis of the as-cast and annealed $V_{0.5}Si_{0.1}Ti_{0.4}$ alloy.

Figure 2(a) shows the SEM images of the as-cast and annealed $V_{0.5}Si_{0.1}Ti_{0.4}$ alloys after etching the polished surface. The elemental analysis of a portion of the as-cast and annealed $V_{0.5}Si_{0.1}Ti_{0.4}$ alloys are shown in the fig. 2(b). The light grey region is rich in V and Ti, indicating the presence of the β-V-Ti phase with 6 at.% Si. The black patches in the boundary region of the β-V-Ti phase are rich in Ti and Si, indicating the presence of the $Ti_5Si_3$ phase. Comparing the SEM images of the both samples, it becomes evident that in the as-cast sample, there was a eutectic structure present around the β-phase, connecting it throughout. However, after annealing, this eutectic structure disappeared, and the β-phase grains are completely isolated from each other.

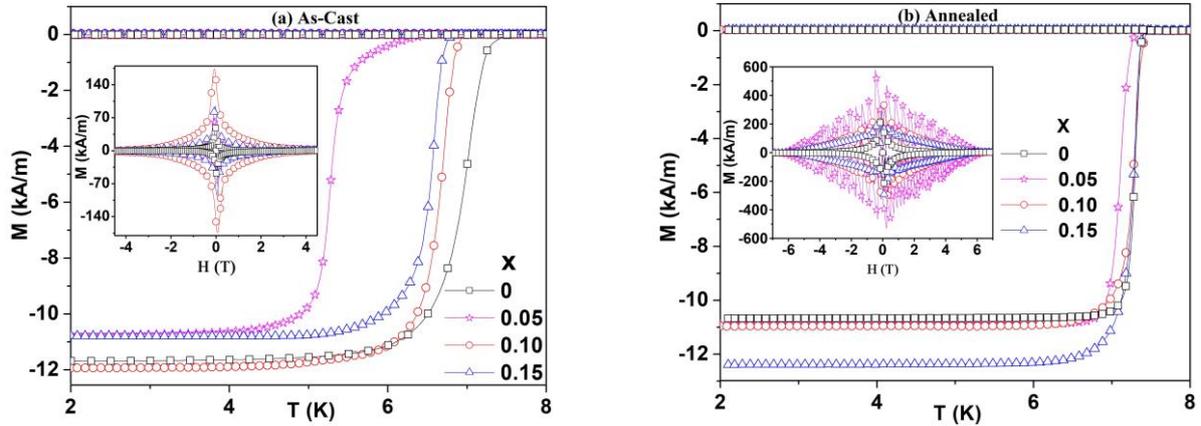

**FIGURE 3.** The low field (10 mT) temperature dependence of magnetization of the (a) as-cast and (b) annealed $V_{0.6-x}Si_xTi_{0.4}$ alloys. Insets to (a) and (b) display the field dependence of magnetization for the as-cast and annealed alloys at 4 K respectively.

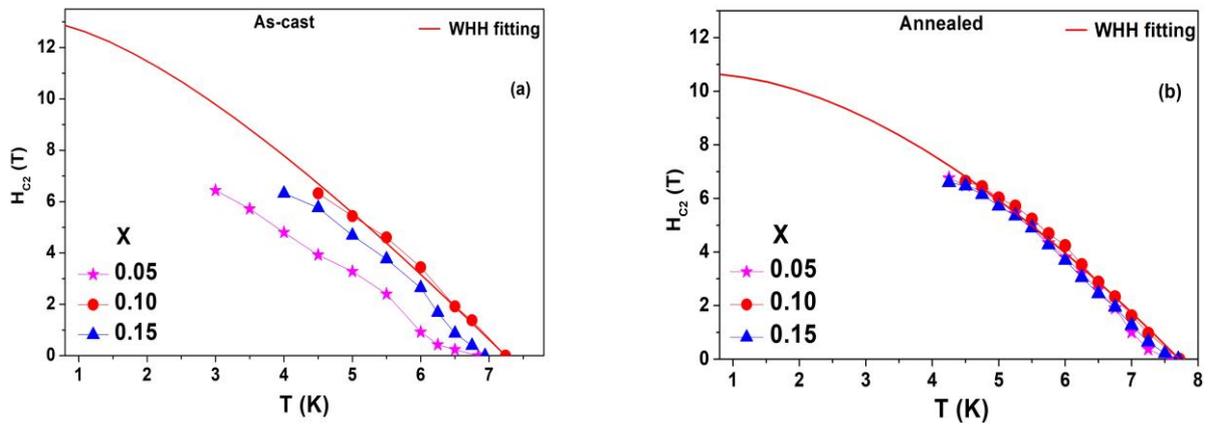

**FIGURE 4.** The temperature dependence of upper critical field ($H_{C2}$) of the (a) as-cast and (b) annealed $V_{0.6-x}Si_xTi_{0.4}$ alloys. The solid symbols are the experimental data and the solid line is the fit using WHH formalism for the x = 0.10 sample.

Figure 3 illustrates the M(T) behavior of (a) as-cast and (b) annealed $V_{0.6-x}Si_xTi_{0.4}$ alloy measured in the temperature range 2- 12 K for an applied field of 10 mT. For each of these samples, the M(T) measurements were performed using three different modes: (i) cooling down to 2 K in zero field from T > $T_C$, applying the field of measurement at 2 K and doing the measurements during warming up (ZFC), (ii) cooling down to 2 K from T > $T_C$ while doing measurements in the same magnetic field (FCC), and (iii) the measurements continued during the subsequent warming up (FCW). The $T_C$ is estimated as the temperature at which the magnetization starts to decrease with decreasing temperature and attains a negative value. The as-cast $V_{0.6}Ti_{0.4}$ alloy has a $T_C$ of 7.62 K. The $T_C$ reduces to 6.84 K for the x = 0.05 alloy. This large drop in $T_C$ is due to the 6 at.% solubility of Si in $V_{0.6}Ti_{0.4}$ alloy. The $T_C$ increases to 7.05 K for the x = 0.10 alloy and then decreases slightly to 6.9 K for x = 0.15. This increase in $T_C$ is due to the enrichment of V in the β-V-Ti phase. After annealing, the $T_C$ increases for all the alloys due to the V enrichment in the β matrix. Except for the x = 0.05 alloy, the $T_C$ of the other annealed alloys is nearly same (7.7 K). In addition, it can also be seen from the fig. 3 that transition becomes sharper after annealing which indicates the reduction of disorder in the β-phase.

The insets to Fig. 3 depict the M(H) curves for the (a) as-cast and (b) annealed $V_{0.6-x}Si_xTi_{0.4}$ alloys measured at 4 K. For the as-cast alloys, addition of Si to the $V_{0.6}Ti_{0.4}$ alloy increases the irreversibility both in magnitude and field range up to x = 0.10. However, for x = 0.15, the hysteresis diminishes in both magnitude and field range. In the case of annealed alloys, on the other hand, the maximum hysteresis is observed for x = 0.05, and as the Si content increases further, the hysteresis gradually decreases.

Figure 4 shows the temperature dependence of upper critical field ($H_{C2}$) of the (a) as-cast and (b) annealed $V_{0.6-x}Si_xTi_{0.4}$ alloys. In the as-cast alloys Si addition increases the $H_{C2}$ due to the increased disorder. For x = 0.05 alloy, the $H_{C2}$ is quite low over a temperature range below $T_C$ indicating the filamentary type superconductivity in this alloy due to the solubility of Si in the β matrix. The $H_{C2}(T)$ is nearly same for all the annealed alloys. The solid lines are the fit to the $H_{C2}(T)$ using Werthamer–Helfand–Hohenberg (WHH) formalism [13]. We see that the $H_{C2}(0)$ decreases after annealing for the x = 0.10 alloy. The coherence length in the limit of absolute zero ($\xi(0)$) is estimated as $\xi^2(0) = 2.07 \times 10^{-15}/2\pi H_{C2}$. The $\xi(0)$ = 4.98 nm for the as-cast x = 0.10 alloy which increases to 5.5 nm after annealing. This is in line with the increase in the $T_C$ after annealing.

## CONCLUSION

In conclusion, it is shown that when Si is added in $V_{1-x}Ti_x$ alloys, it forms in a two-phase microstructure for Si contents higher than 6 at.%: a main β-phase and a secondary hexagonal $Ti_5Si_3$ phase. Moreover, polycrystalline as-cast $V_{0.6-x}Si_xTi_{0.4}$ alloys show a eutectic microstructure around the β-phase which disappears after annealing. Additionally, while the $T_C$ is found to change non-monotonically with increasing x in the as-cast $V_{0.6-x}Si_xTi_{0.4}$ alloys, the $T_C$ increases after annealing for all the alloys. On the other hand, the $H_{C2}(0)$ decreases and the coherence length increases after annealing in the x = 0.10 alloy. Analysis of present results show that the present findings can be attributed to the precipitation of the $Ti_5Si_3$ phase, which subsequently causes V enrichment in the β-phase.

## ACKNOWLEDGMENTS

Asi Khandelwal is thankful to HBNI for the financial support.

## REFERENCES

1. M. Tai, K. Inoue, A. Kikuchi, T. Takeuchi, T. Kiyoshi and Y. Hishinuma, IEEE Trans. Appl. Supercond., 17, 2542–2545, (2007).

2. T. Takeuchi, H. Takigawa, M. Nakagawa, N. Banno, K. Inoue, Y. Iijima and A. Kikuchi, Supercond. Sci. Technol. 21, 025004, (2008).

3. H. Zhang, J.P. Lin, Y.F. Liang, S. Xu, Y. Xu, S.-L. Shang, Z.-K. Liu, Intermetallics 115, 106609, (2019).

4. M. Regenberg, G. Hasemann, C. Müller, and M. Krüger, IOP Conf. Ser.: Mater. Sci. Eng. 882, 012014 (2020).

5. B. B. de Lima-Kühn, A. A. A. P. da Silva, P. A. Suzuki, G. C. Coelho, and C. A. Nunes, Mater. Res. 19, 1122 (2016).

6. M. Fiore, F. Beneduce, and C. R. de F. Azevedo, Technol. Metal. Mater. Miner. (Sao Paulo), 13, 91 (2016).

7. M. Enomoto, J. Ph. Equilibria, 13, 201 (1992).

8. B. Hu, B. Yao, J. Wang, Y. Liu, C. Wang, Y. Du, and H. Yin, Intermetallics, 118, 106701, (2020).

9. S. Komjathy, J. Less-Com. Met. 3, 468-488, (1961).

10. M. Touminen, G. W. Franti, and D. A. Koss, Metal. Trans. 8A, 457 (1977).

11. A. Inoue, C. Suryanarayana, T. Masumoto, and A. Hoshi, Mater. Sci. Eng. 47, 59 (1981).

12. G. Nolze, "A mixture between crystal structure visualizer, simulation and refinement tool," in: Powder Diffraction: Proc. of the II Internat. School on Powder Diffraction, Kolkata (2002), pp. 146–155.

13. N.R. Werthamer, E. Helfand and P.C. Hohenberg, Phys. Rev. 147, 295 (1966).